\documentclass[prl,twocolumn,letter,superscriptaddress]{revtex4-1}
\usepackage{graphicx,psfrag}
\usepackage{bm}
\usepackage{mathbbol,verbatim}
\usepackage{slashed}
\usepackage{multirow}
\usepackage{graphics}
\usepackage{color,ulem, amsmath, amssymb}
\usepackage[colorlinks=true,linkcolor=blue,citecolor=blue]{hyperref}%

\definecolor{greeen}{rgb}{0.03,0.84,0.13}
\definecolor{test}{rgb}{0.03,0.74,0.33}
\definecolor{viol}{rgb}{0.44,0,0.94}
\definecolor{or}{rgb}{0.95,0.65,0}

\newcommand{\blue}{\textcolor{blue}}

\begin{document}

\title{Lepton Flavor Violation Induced by a Neutral Scalar at Future Lepton Colliders}

\author{P. S. Bhupal Dev}
\affiliation{Department of Physics and McDonnell Center for the Space Sciences,  Washington University, St. Louis, MO 63130, USA}

\author{Rabindra N. Mohapatra}
\affiliation{Maryland Center for Fundamental Physics, Department of Physics, University of Maryland, College Park, MD 20742, USA}

\author{Yongchao Zhang}
\affiliation{Department of Physics and McDonnell Center for the Space Sciences,  Washington University, St. Louis, MO 63130, USA}

\begin{abstract}
  Many new physics scenarios beyond the Standard Model  often necessitate the existence of a (light) neutral scalar $H$, which might couple to the charged leptons in a flavor violating way, while evading all existing constraints. We show that such scalars could be effectively produced at future lepton colliders, either on-shell or off-shell depending on their mass, and induce lepton flavor violating (LFV) signals, i.e. $e^+ e^- \to \ell_\alpha^\pm \ell_\beta^\mp (+H)$ with $\alpha\neq \beta$. We find that a large parameter space of the scalar mass and the LFV couplings can be probed, well beyond the current low-energy constraints in the lepton sector. In particular, a scalar-loop induced explanation of the longstanding muon $g-2$ anomaly can be directly tested in the on-shell mode.
\end{abstract}

\maketitle

{\bf Introduction.--}
The observation of neutrino oscillations~\cite{PDG} suggests that the
lepton family numbers are violated. It also calls for an extension of the Standard Model (SM) to include neutrino mass terms, which necessarily induce charged lepton flavor violation (cLFV). In the minimal extension of the SM with Dirac neutrinos, cLFV rates are highly suppressed due to small neutrino masses (compared to the electroweak scale). This makes the experimental searches for cLFV all the more interesting, because any observable effect must come from physics beyond the minimally extended SM related to the origin of neutrino mass.

There are various theoretical models of new physics which lead to cLFV effects at an observable level~\cite{deGouvea:2013zba, Lindner:2016bgg}. They generally involve extending the Higgs sector, which allows flavor-violating Yukawa couplings of new neutral scalars beyond the SM.
In particular, if any of the new neutral scalars (call it $H$) is (almost) hadrophobic, it could remain sufficiently light and contribute sizably to cLFV, while easily evading the direct searches at hadron colliders, as well as the low-energy quark flavor constraints, such as the rare flavor-changing decays and oscillations of $K$ and $B$ mesons. Some well-motivated examples include supersymmetric models with leptonic $R$-parity violation~\cite{susy}, left-right symmetric models~\cite{Dev:2016vle}, mirror models~\cite{mirror}, and two-Higgs doublet models~\cite{2HDM}, where the cLFV coupling might arise at tree or loop level~\cite{supp}.

In this letter, we show that such scenarios of neutral scalar-induced cLFV can be effectively probed in a model-independent way at future lepton colliders, such as the Circular Electron-Positron Collider (CEPC)~\cite{CEPC-SPPCStudyGroup:2015csa}, International Linear Collider (ILC)~\cite{Baer:2013cma}, Future Circular Collider (FCC-ee)~\cite{Gomez-Ceballos:2013zzn} and Compact Linear Collider (CLIC)~\cite{Battaglia:2004mw}.
Compared to the hadron colliders, the lepton colliders are generally very ``clean'' and the SM processes therein are well understood, which render them primary facilities to search for new physics  via the cLFV signals $e^+ e^- \to \ell_\alpha^\pm \ell_\beta^\mp + X$ (with ${\alpha,\,\beta} = e,\, \mu,\, \tau$ and $\alpha \neq \beta$). Previous studies of LFV at lepton colliders have either been performed in the framework of effective four-fermion couplings~\cite{Kabachenko:1997aw, Ferreira:2006dg} or in the context of flavor-violating SM Higgs decays~\cite{Banerjee:2016foh} and tau decays~\cite{Hays:2017ekz} or with doubly-charged scalars~\cite{Rodejohann:2010bv}. Here we include both on and off-shell production of the new neutral scalar $H$ (including resonance) at lepton colliders, which enables us to derive the LFV sensitivity as a function of the mass $m_H$ for a direct comparison with the current bounds from low-energy experiments. Moreover, for $m_H$ small compared to the center-of-mass energy, the effective theory approximation does not work.

Without loss of generality, we can write the effective Yukawa couplings of $H$ to the charged leptons as
\begin{eqnarray}
\label{eqn:Yukawa}
{\cal L}_Y \ = \ h_{\alpha \beta}  \bar{\ell}_{\alpha,\,L} H \ell_{\beta,\, R} ~+~ {\rm H.c.} \, .
\end{eqnarray}
Here for simplicity we have assumed the couplings are all real and chirality-independent and thus symmetric. The scalar $H$ may or may not be responsible for symmetry breaking and/or mass generation of other particles in realistic models, where it could be part of a singlet, doublet or triplet scalar field.  We assume that it is CP even and its mixing with and/or coupling to the SM Higgs is small. If the scalar is CP-odd, the limits and prospects derived in this Letter would not change significantly. 
Even though there are all varieties of stringent low-energy cLFV constraints, such as $\ell_\alpha\to \ell_\beta \gamma$, $\ell_\alpha\to 3\ell_\beta, 2\ell_\beta\ell_\gamma$~\cite{PDG}, only a few of them are {\it directly} relevant to the LFV prospects discussed below. With an ab$^{-1}$ level of integrated luminosity,  a large parameter space of $m_{H}$ and $h_{\alpha\beta}$ could be probed, well beyond the current cLFV constraints and complementary to the projected low-energy constraints from future experiments at the intensity frontier~\cite{Calibbi:2017uvl}. In addition, the Lagrangian~\eqref{eqn:Yukawa} also gives rise to a one-loop contribution to the lepton anomalous magnetic moment. In particular, the longstanding muon $g-2$ discrepancy~\cite{PDG} could also be tested {\it directly} at lepton colliders.

{\bf On-shell LFV.--}
If kinematically allowed, the neutral scalar $H$ can be directly produced at lepton colliders, in association with a pair of flavor-changing leptons through the couplings in Eq.~(\ref{eqn:Yukawa}), i.e.
$e^+ e^- \to \ell^\pm_\alpha \ell^\mp_\beta H$ (with $\alpha \neq \beta$),
as shown in Fig.~\ref{fig:diagram1} (top panel).
\begin{figure}[t!]
  \includegraphics[width=6cm]{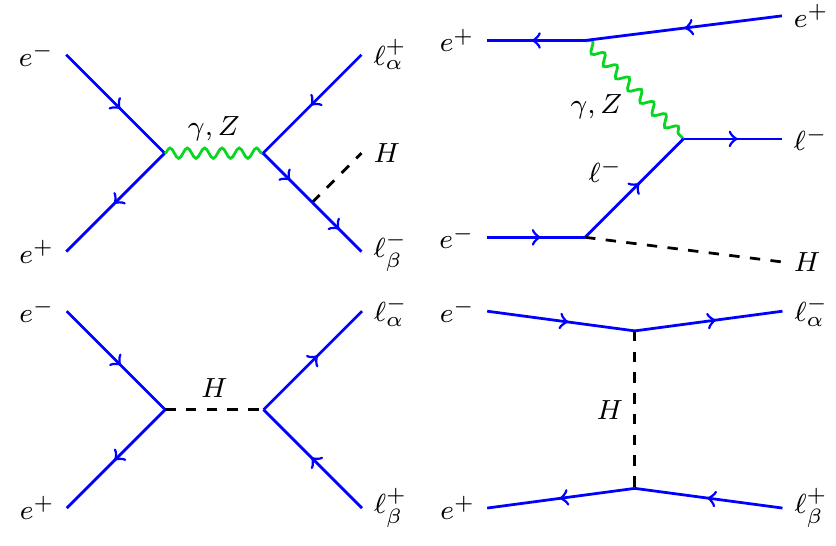}
  \vspace{-5pt}
  \caption{Examples of LFV signal from on-shell, off-shell and resonant neutral scalar production at $e^+e^-$ collider.}
  \label{fig:diagram1}
  \vspace{-8pt}
\end{figure}

{\bf \boldmath$e-\mu$ coupling:}
Here for simplicity we assume the other two $h_{e\tau}$ and $h_{\mu\tau}$ are vanishing. It should be emphasized that the amplitudes in Fig.~\ref{fig:diagram1} depend only on the LFV couplings $h_{\alpha\beta}$ (here $\alpha\beta = e\mu$), and thus could be easily made to satisfy the rare lepton decay constraints, such as $\mu \to eee$ and $\mu \to e \gamma$, which depend on the product $|h_{ee}^\dagger h_{e\mu}|$. Similarly, with vanishing or suppressed couplings to the quark sector, the $\mu - e$ conversion limits are irrelevant. Finally, for real Yukawa couplings, we do not either have any limits from electric dipole moment. Thus we are left only with the following constraints (summarized in Table~\ref{tab:limits}):

({\it i}) {\it Muonium-antimuonium oscillation:} This could occur in both $s$ and $t$-channels~\cite{supp}. The oscillation probability ${\cal P} \propto |h_{e\mu}|^4/m_H^4$. The MACS experiment~\cite{Willmann:1998gd} could then exclude a large parameter space, as shown in the left panel of Fig.~\ref{fig:p1}.

\begin{table}[b]
  \vspace{-13pt}
  \centering
  \footnotesize
  \caption[]{Current experimental constraints on the LFV couplings. The $\Delta a_{e,\,\mu}$ constraints have an additional logarithmic dependence on the scalar mass~\cite{supp}. }
  \vspace{-5pt}
  \label{tab:limits}
  \begin{tabular}[t]{c|c}
  \hline\hline
  process & 
  constraints $\times ({\rm GeV}/m_H)^{2}$ \\ \hline
  muonium oscillation & 
  $|h_{e\mu}|^2 < 1.0 \times 10^{-7}$ \\ \hline

  \multirow{2}{*}{$\Delta a_e$} & 
  $|h_{e\mu}|^2 < 6.2 \times 10^{-8}$ \\
  & $|h_{e\tau}|^2 < 6.9 \times 10^{-9}$ \\ \hline

  $\Delta a_\mu$ & 
  $|h_{\mu\tau}|^2 < 4.4 \times 10^{-7}$ \\ \hline

  $ee \to \mu\mu$ & 
  $|h_{e\mu}|^2 < 1.6 \times 10^{-7}$ \\
  $ee \to \tau\tau$ & 
  $|h_{e\tau}|^2 < 1.0 \times 10^{-7}$ \\ \hline

  $\mu^- \to e^- e^+ e^-$ & 
  $|h_{ee}^\dagger h_{e\mu}| < 6.6 \times 10^{-11}$ \\ \hline

  $\tau^- \to e^- e^+ e^-$ & 
  $|h_{ee}^\dagger h_{e\tau}| < 2.6 \times 10^{-8}$ \\
  $\tau^- \to \mu^- e^+ e^-$ & 
  $|h_{ee}^\dagger h_{\mu\tau}| < 1.5 \times 10^{-8}$ \\
  $\tau^- \to \mu^+ e^- e^-$ & 
  $|h_{e\mu}^\dagger h_{e\tau}| < 1.9 \times 10^{-8}$ \\
  $\tau^- \to e^- \gamma$ & 
  $|h_{ee}^\dagger h_{e\tau}| < 1.0 \times 10^{-6}$ \\
  $\tau^- \to \mu^- \gamma$ & 
  $|h_{e\mu}^\dagger h_{e\tau}| < 1.2 \times 10^{-6}$ \\ \hline

  \multirow{2}{*}{$\Delta a_e$} & 
  $|h_{ee}^\dagger h_{e\tau}| < 1.1 \times 10^{-7}$ \\
  & $|h_{e\mu}^\dagger h_{e\tau}| < 1.0 \times 10^{-8}$ \\ \hline

  $ee \to ee,\, \tau\tau$ & 
  $|h^\dagger_{ee} h_{e\tau}| < 1.4 \times 10^{-7}$ \\
  $ee \to \mu\mu,\, \tau\tau$ & 
  $|h^\dagger_{e\mu} h_{e\tau}| < 1.3 \times 10^{-7}$ \\
  \hline\hline
  \end{tabular}
\end{table}

({\it ii}) {$(g-2)_e$:} The anomalous magnetic moment of electron $a_e$ receives a contribution from the $H - \mu$ loop~\cite{supp}.
As a result of the precise measurement of $a_e$~\cite{Mohr:2015ccw}, 
the constraint on $h_{e\mu}$ is comparable to that from muonium oscillation, as shown in the left panel of Fig.~\ref{fig:p1}. 
To explain the longstanding theoretical and experimental discrepancy of the muon $g-2$, i.e. $\Delta a_\mu = (2.87 \pm 0.80) \times 10^{-9}$~\cite{PDG}, the LFV coupling $h_{e\mu}$ is required to be larger, as shown in the left panel of Fig.~\ref{fig:p1} (by the brown line and the green, yellow bands corresponding respectively to the central value and the $1\sigma$, $2\sigma$ ranges), which is already excluded by the $(g-2)_e$ data. See however the $\mu\tau$ sector in the right panel of Fig.~\ref{fig:p1} for an explanation of  $(g-2)_\mu$ in this setup. 

({\it iii}) {$e^+e^-\to \mu^+\mu^-$:} A $t$-channel $H$ could mediate the scattering $e^+ e^- \to \mu^+ \mu^-$, which interferes with the SM diagrams in the $s$-channel~\cite{Hou:1995dg}. Both the total cross section and differential distributions would be modified by the presence of $H$, depending on its mass and the coupling $h_{e\mu}$. In the heavy $H$ limit, the LEP data exclude an effective cutoff scale $\Lambda \simeq m_{H}/h_{e\mu}$~\cite{Abdallah:2005ph}. When $H$ is lighter than the center-of-mass energy $\sqrt{s}$, the limits on $\Lambda$ do not apply, and we consider the $H$ propagator: $(q^2 - m_{H}^2)^{-1} = (-s \cos\theta/2 - m^2_{H})^{-1}$. For simplicity we take an average over the scattering angle $\langle \cos\theta \rangle \simeq 1/2$ to interpret the LEP constraints. Then in the limit of $m_{H} \ll \sqrt{s}$, the propagator is dominated by the $q^2$ term, and the $ee \to \mu\mu$ limit in Fig.~\ref{fig:p1} approaches a constant, as expected.

\begin{figure}[b!]
  \vspace{-5pt}
  \centering
  \includegraphics[width=0.23\textwidth]{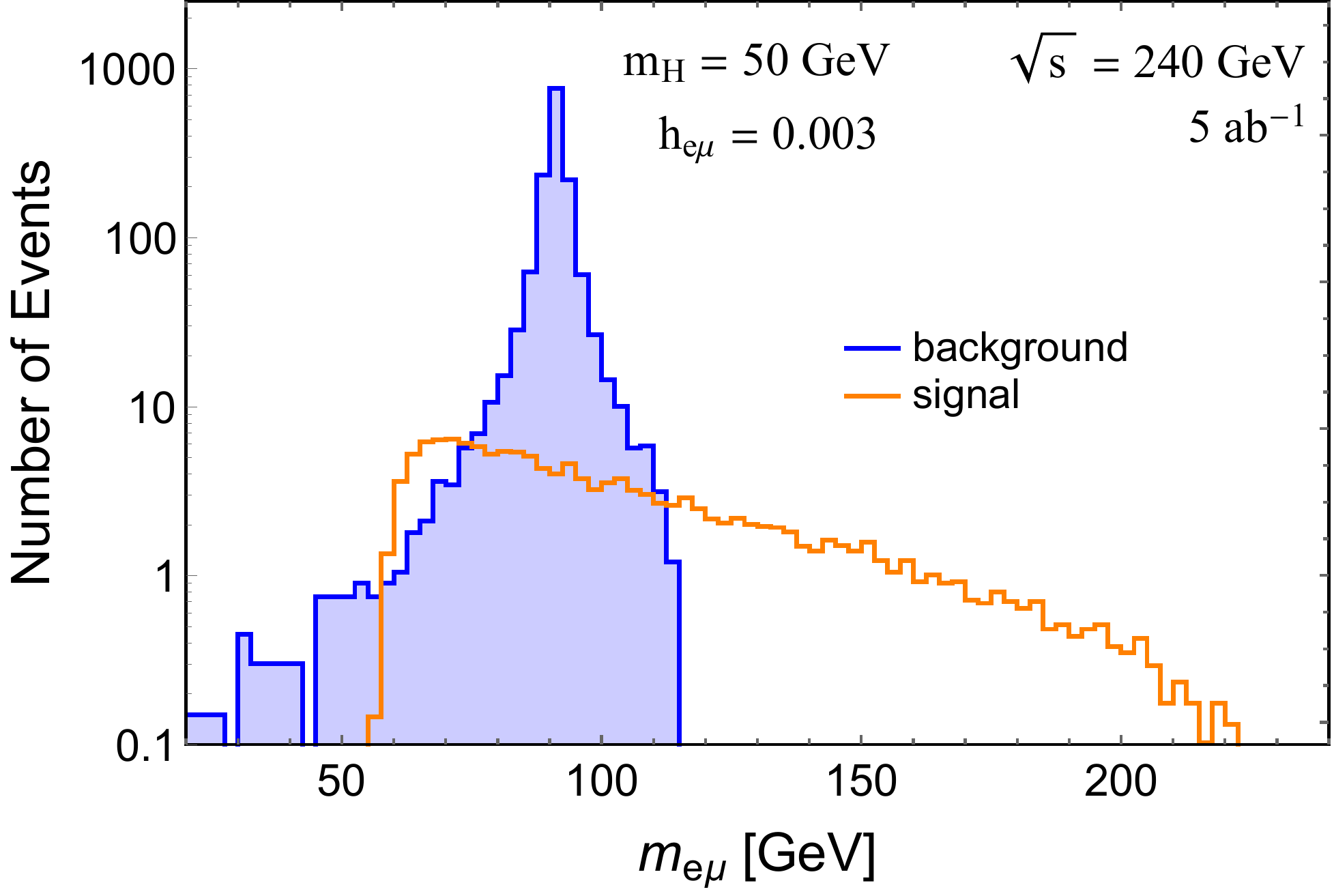} \hspace{0.1cm}
  \includegraphics[width=0.235\textwidth]{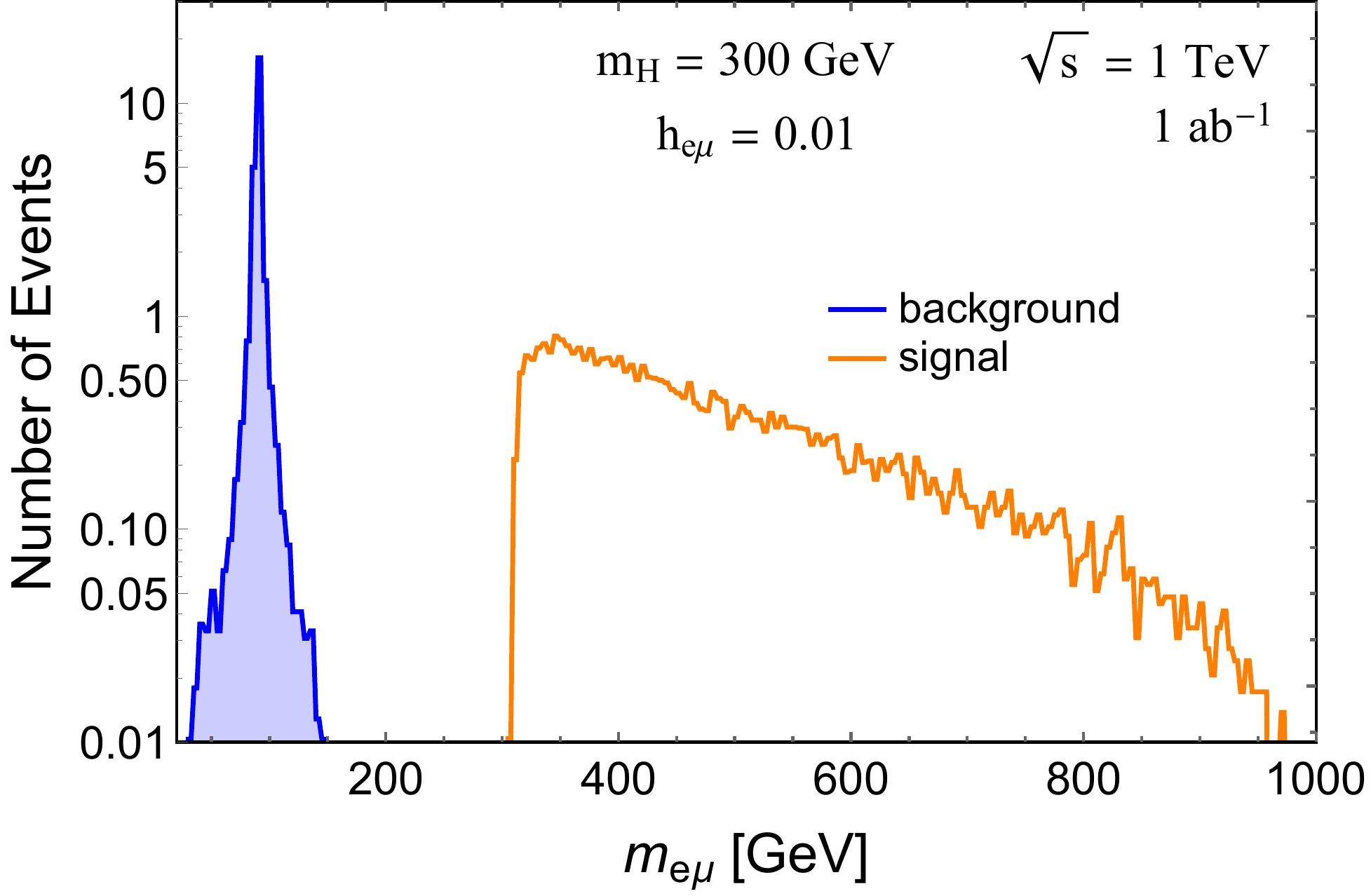} \vspace{-15pt}
  \caption{$m_{e\mu}$ invariant mass distributions for the SM background $e^+ e^- \to Zh$ and the signal $e^+ e^- \to e^\pm \mu^\mp H$ at CEPC (left) and ILC (right).  }
  \label{fig:example}
\end{figure}

\begin{figure*}[htb]
  \centering
  \includegraphics[width=0.31\textwidth]{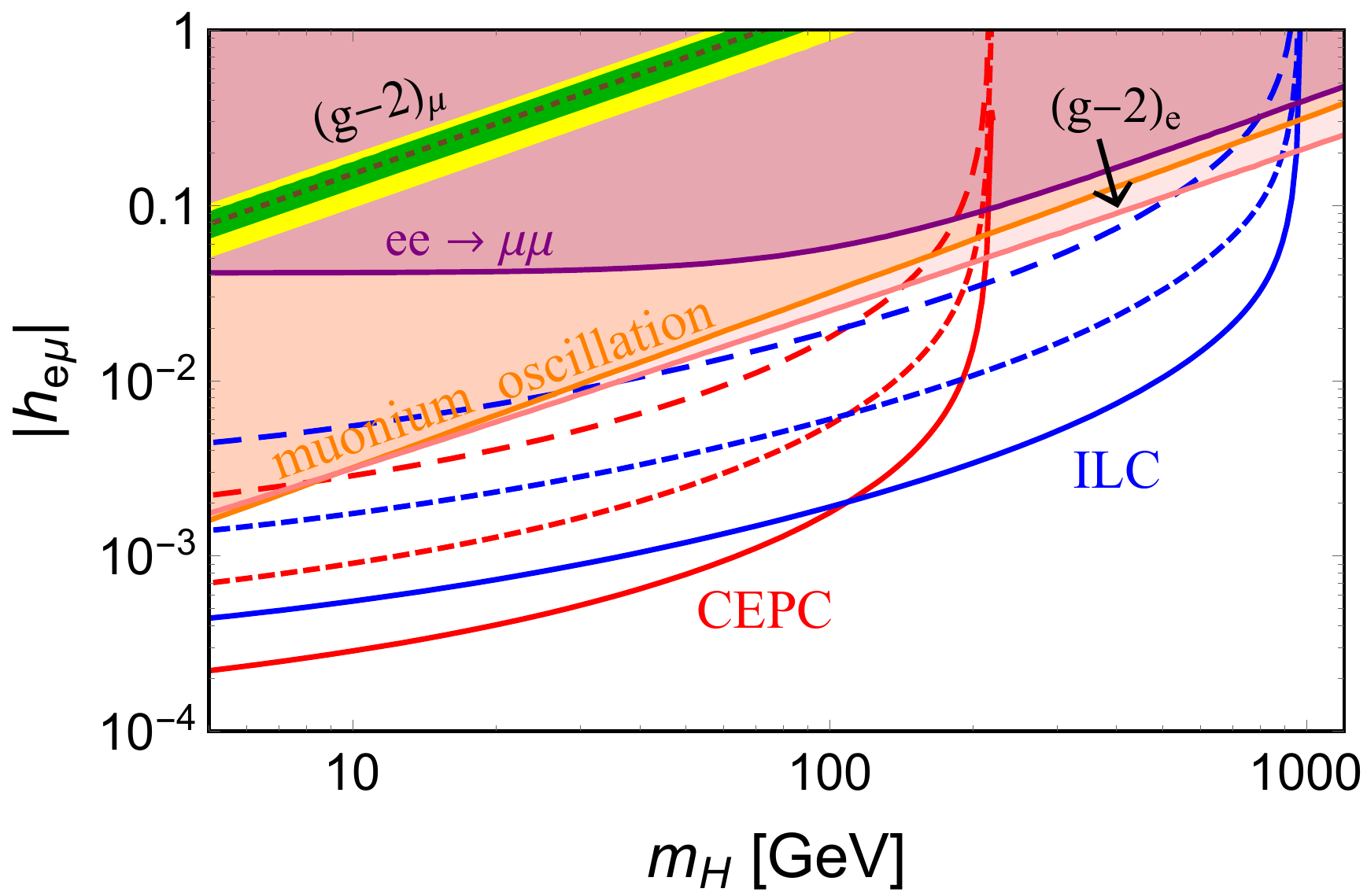}
  \includegraphics[width=0.31\textwidth]{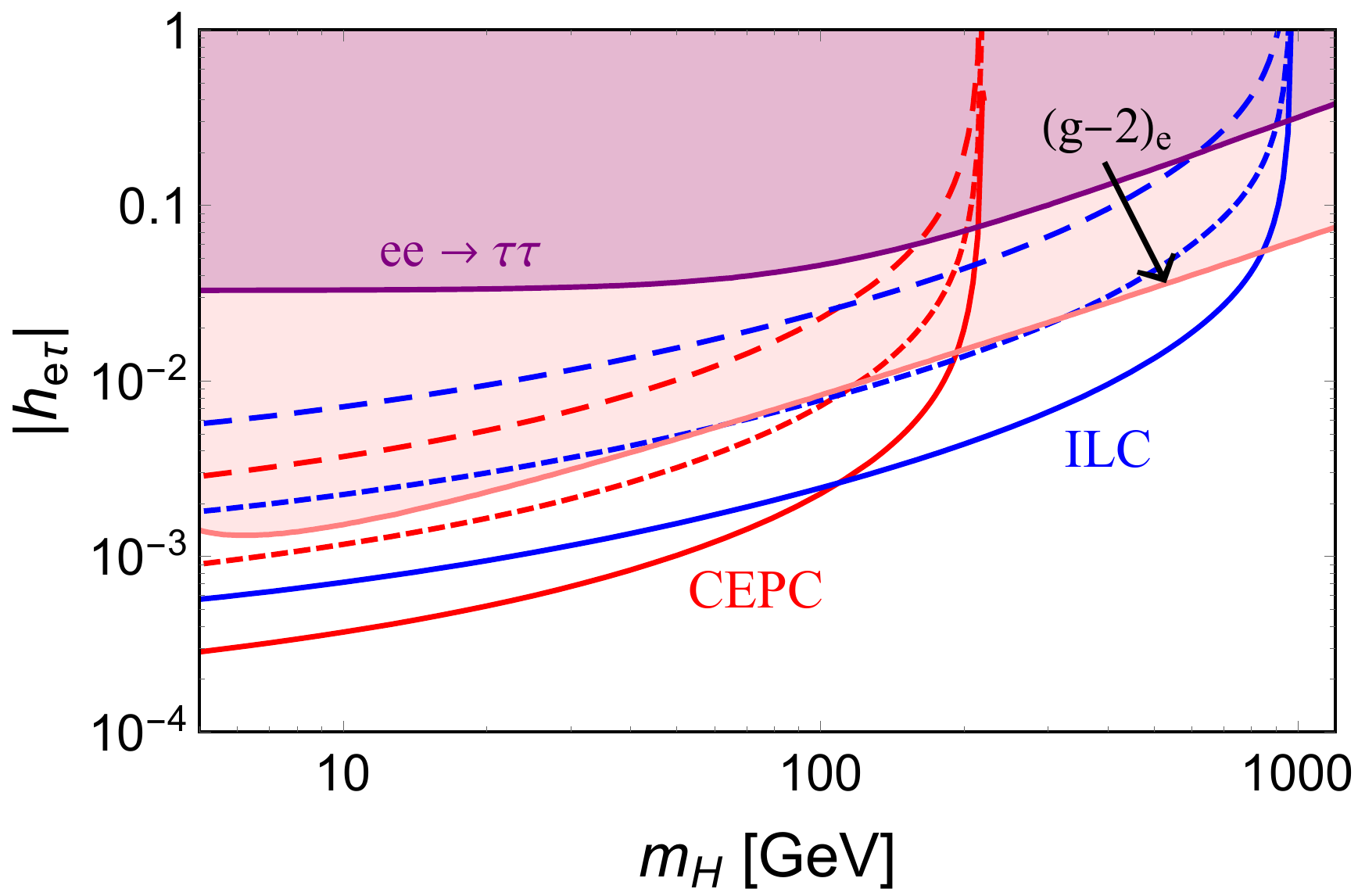}
  \includegraphics[width=0.31\textwidth]{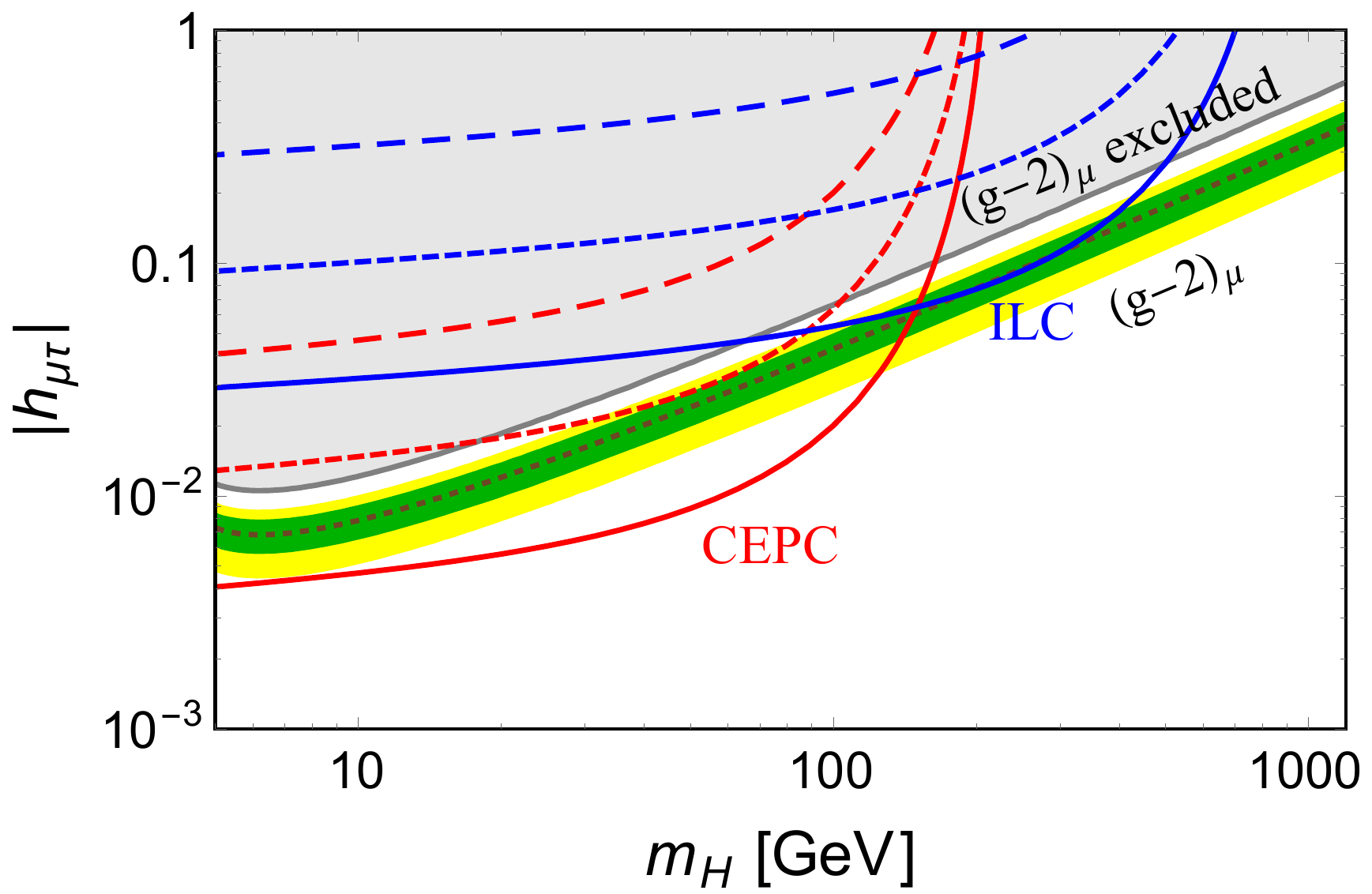}
  \vspace{-5pt}
  \caption{Prospects of probing LFV couplings $h_{\alpha\beta}$ ($\alpha \neq \beta$) from searches of $e^+ e^- \to \ell_\alpha^\pm \ell_\beta^\mp H$ at CEPC (red, $\sqrt{s} = 240$ GeV and ${\cal L} = 5$ ab$^{-1}$) and ILC (blue, 1 TeV and 1 ab$^{-1}$). Here we have assumed 10 LFV signal events and a BR of 1\% (long-dashed) 10\% (short-dashed) or 100\% (solid) from $H$ decay to be visible.
  In the left panel, the region shaded in orange, pink and purple and respectively excluded by muonium oscillation, $(g-2)_e$ and $ee \to \mu\mu$ data; in the middle panel, the pink and purple regions are excluded by $(g-2)_e$ and $ee \to \tau\tau$ data; in the right panel the gray region is disfavored by the $(g-2)_\mu$ data at the $5\sigma$ confidence level. In the left and right panels, the brown line could fit the central value of $\Delta a_\mu$, and the green and yellow bands cover the $1\sigma$ and $2\sigma$ ranges of $\Delta a_\mu$.  }
  \vspace{-5pt}
  \label{fig:p1}
\end{figure*}

To be specific, we consider two benchmark configurations for future lepton colliders: i.e. the CEPC~\cite{CEPC-SPPCStudyGroup:2015csa} and ILC~\cite{Baer:2013cma}, with the center-of-mass energies $\sqrt{s}$, integrated luminosities, and the nominal cuts on the leptons $\ell$ (implemented by using {\tt CalcHEP}~\cite{Belyaev:2012qa}) summarized in Table~\ref{tab:collider}. The total cross sections $\sigma (ee \to \ell_\alpha \ell_\beta (+H))$ in the light $H$ limit (and $m_H = 100$ GeV for the on-shell production) are also presented in the table, with a conservative efficiency of 60\% for the $\tau$ lepton~\cite{Baer:2013cma}.
The systematic uncertainties such as initial state radiation, beamstrahlung, and the electron and muon efficiencies lead only up to a few percent correction to the total cross sections~\cite{CEPC-SPPCStudyGroup:2015csa,Baer:2013cma}.

\begin{table}[t]
  \vspace{-5pt}
  \centering
  \caption[]{Benchmark configurations of future lepton colliders CEPC and ILC and the expected total cross sections of the on-shell and off-shell production of $H$, up to the LFV couplings squared, in the light $H$ limit. The values in parentheses are for $m_H = 100$ GeV.}
  \label{tab:collider}
  \vspace{-5pt}
  \begin{tabular}[t]{c|c|c}
  \hline\hline
  collider & CEPC & ILC \\ \hline
  $\sqrt{s}$ & 240 GeV & 1 TeV \\
  luminosity & 5 ab$^{-1}$ & 1 ab$^{-1}$ \\ \hline
  cuts & \multicolumn{2}{c}{ $p_T (\ell) > 10$ GeV, $|\eta(\ell)| < 2.5$ } \\ \hline
  $\sigma (e\mu + H_3) / |h_{e\mu}|^2$ &
  $8.9 \times 10^4 \, (390)$ fb &  $1.1 \times 10^5 \, (2800)$ fb \\
  $\sigma (e\tau + H_3) / |h_{e\tau}|^2$ &
  $5.3 \times 10^4 \, (650)$ fb &  $6.6 \times 10^4 \, (1700)$ fb \\
  $\sigma (\mu\tau + H_3) / |h_{\mu\tau}|^2$ & $2100 \, (5.0)$ fb &  $5700 \, (3.5)$ fb \\ \hline

  $\sigma (e\tau) / |h^\dagger_{ee} h_{e\tau}|^2$ &
  $4.8 \times 10^5$ fb & $2.8 \times 10^4$ fb \\
  $\sigma (\mu\tau) / |h^\dagger_{ee} h_{\mu\tau}|^2$ &
  $1.6 \times 10^5$ fb & $9300$ fb \\
  $\sigma (\mu\tau) / |h^\dagger_{e\mu} h_{e\tau}|^2$ &
  $1.6 \times 10^5$ fb & $9300$ fb \\
  \hline\hline
  \end{tabular}
  \vspace{-5pt}
\end{table}

The SM background is dominated by particle mis-identification from the Higgsstrahlung process $e^+ e^- \to Zh$ with one of the $e$ ($\mu$) from $Z$ decay mis-identified as $\mu$ ($e$)~\cite{Yu:2017mpx} (see also~\cite{Hammad:2016bng}).
The invariant mass $m_{e\mu}$ distributions from the on-shell production of $e\mu H$ can be easily distinguished from the backgrounds, as exemplified in Fig.~\ref{fig:example}, with $m_H = 50$ GeV and $h_{e\mu} = 0.003$ at CEPC, and with  $m_H = 300$ GeV and $h_{e\mu} = 0.01$ at ILC. Removing the $Z$-resonance peak, the LFV signal is almost background free. Summing all the bins off the $Z$-peak, the signal ($S$) to background ($B$) significance $S/\sqrt{S+B}$ for the examples in Fig.~\ref{fig:example} are respectively 55 and 61.

After being produced, $H$ could decay back into the charged lepton pairs or other SM particles. Reconstructing the $H$ peak from the decay products could improve further the significance of the LFV signals, which are however rather model-dependent. 
To work in a model-independent way, we consider three benchmark values, where 1\%, 10\% or 100\% of the decay products of $H$ are {\it visible} and can be reconstructed. The corresponding LFV prospects are shown in the left panel of Fig.~\ref{fig:p1}, where we have assumed a minimum of 10 signal events at both CEPC and ILC. 
It is clear from Fig.~\ref{fig:p1} that with a BR of $\gtrsim 10\%$, a large region of $m_{H}$ and $|h_{e\mu}|$ can be probed in future lepton colliders, which extends the limits well beyond what is currently available.

{\bf \boldmath$e-\tau$ coupling:}
Turning now to the coupling $h_{e\tau}$, the most stringent limit comes from the electron $g-2$, which is similar to the case of $h_{e\mu}$ except for the enhancement by the $\tau$ mass {[cf. Eq.~(S13)]}, as shown by the pink region in the middle panel of Fig.~\ref{fig:p1}. The LEP $e^+e^- \to \tau^+\tau^-$ limit is slightly stronger than the muon case~\cite{Abdallah:2005ph}, as shown by the shaded purple region in Fig.~\ref{fig:p1}. 
The reconstruction of $\tau$ lepton is more challenging than $\mu$, and thus the prospects of $h_{e\tau}$ are somewhat weaker than $h_{e\mu}$, but there is still ample parameter space to probe at both CEPC and ILC, as long as the effective BR is $\gtrsim 10\%$.

{\bf \boldmath$\mu-\tau$ coupling:} Turning now to the coupling $h_{\mu\tau}$, there are currently no experimental limits, except for the muon $g-2$ discrepancy. This could be explained in presence of $H$ when it couples to muon and tau, as shown by the brown line and the green and yellow bands in the right panel of Fig.~\ref{fig:p1}, while the shaded region is excluded by the current muon $g-2$ data at the 5$\sigma$ level. As $\mu\tau$ can only be produced in $e^+e^-$ collider in the $s$-channel in Fig.~\ref{fig:diagram1}, the production cross section is smaller than those of $e \mu$ and $e\tau$. From {Eq.~(S13)} (with the couplings and lepton masses changed accordingly), the $(g-2)_\mu$ anomaly can be directly tested at CEPC up to a scalar mass of $\simeq 100$ GeV, as shown in Fig.~\ref{fig:p1},  as long as there is a sizable BR of $H$ into visible states. With a larger luminosity being planned~\cite{Gomez-Ceballos:2013zzn}, FCC-ee could do even better.

\begin{figure*}[!t]
  \centering
  \includegraphics[width=0.31\textwidth]{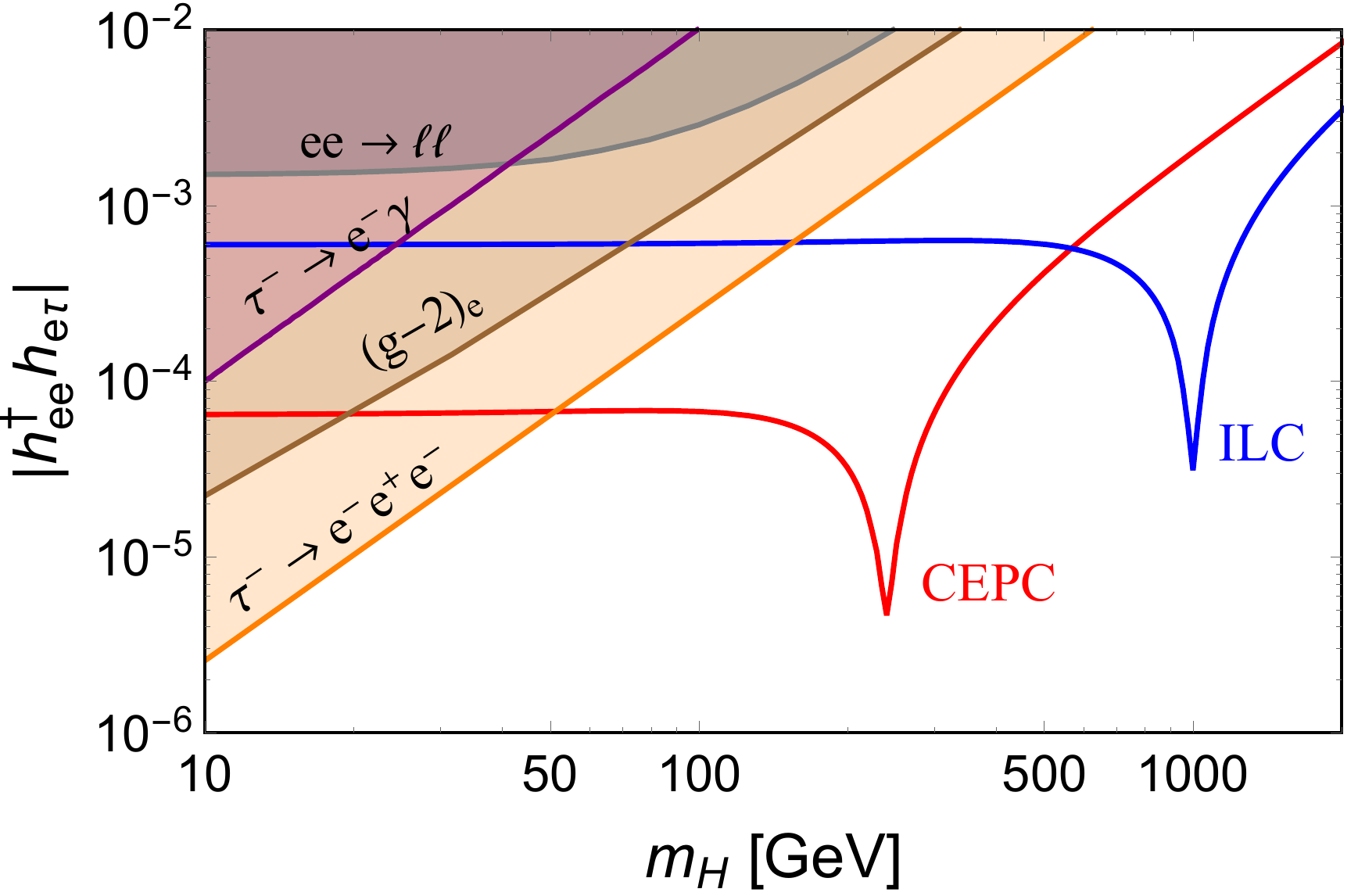}
  \includegraphics[width=0.31\textwidth]{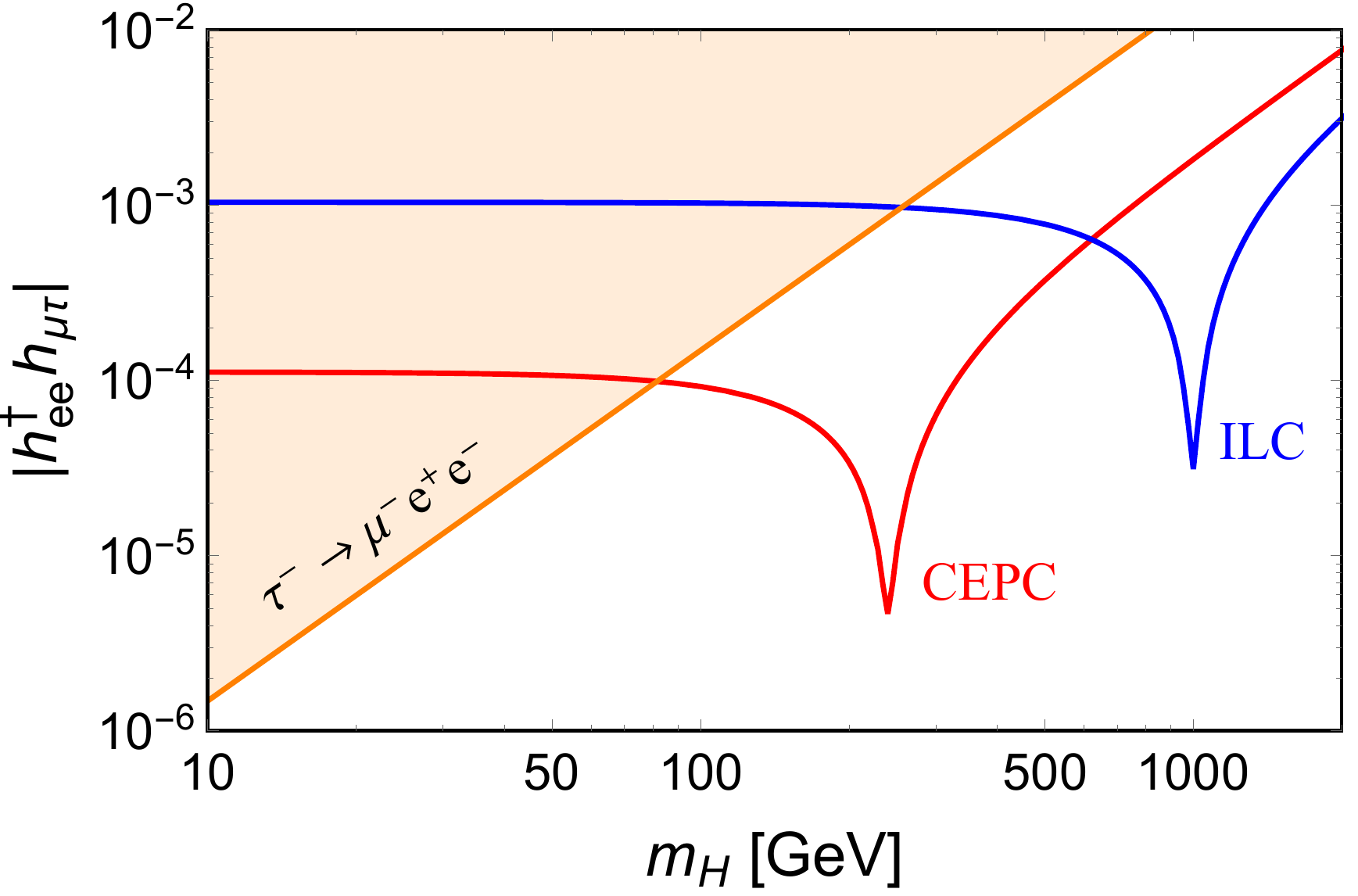}
  \includegraphics[width=0.31\textwidth]{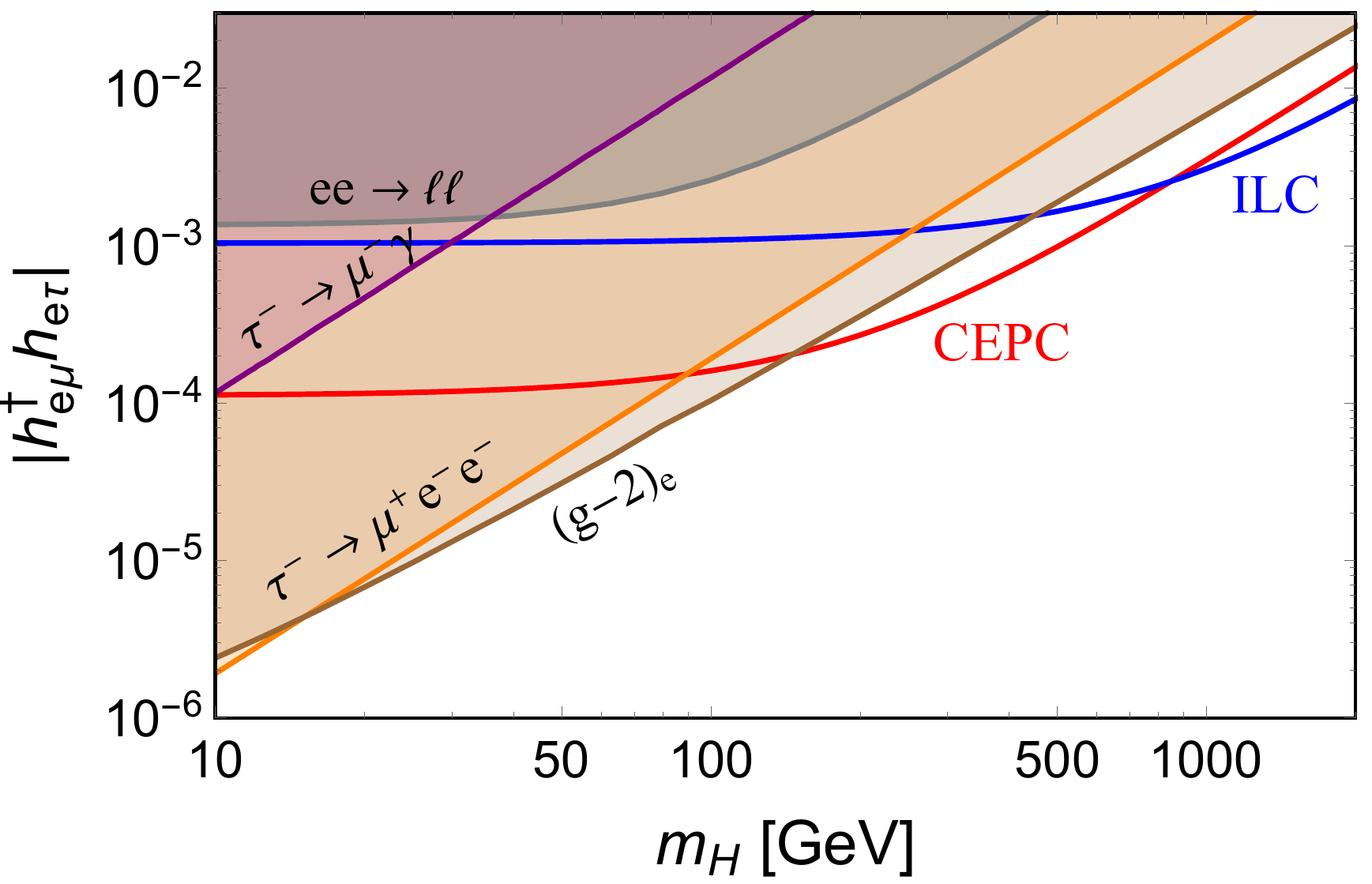}
  \vspace{-5pt}
  \caption{Prospects of $|h^\dagger_{ee} h_{e\tau}|$ (left), $|h^\dagger_{ee} h_{\mu\tau}|$ (middle) and $|h^\dagger_{e\mu} h_{e\tau}|$ (right) from searches of $e^+ e^- \to e^\pm \tau^\mp,\, \mu^\pm \tau^\mp$ at CEPC (red, $\sqrt{s} = 240$ GeV, ${\cal L} = 5$ ab$^{-1}$) and ILC (blue, 1 TeV and 1 ab$^{-1}$). Here we have assumed 10 signal events. Also shown are the constraints from the rare lepton decays, $(g-2)_e$, and $e^+ e^- \to \ell^+ \ell^-$ LEP data (cf. Table~\ref{tab:limits}). }
  \vspace{-5pt}
  \label{fig:p2}
\end{figure*}

{\bf Off-shell (\& resonant) LFV.--}
The LFV signals could also be produced from an off-shell $H$, i.e.
$
e^+ e^- \to \ell^\pm_\alpha \ell^\mp_\beta \,,
$
as shown in Fig.~\ref{fig:diagram1} (bottom panel). This could occur in both the $s$ and $t$ channels; in the $s$-channel $H$ is on-shell if the colliding energy $\sqrt{s} \simeq m_{H}$ (resonance). Different from the on-shell case, the off-shell production amplitudes have a quadratic dependence on the Yukawa couplings (some of them might be flavor conserving), and thus largely complementary to the on-shell LFV  searches.

The amplitude $e^+ e^- \to e^\pm \mu^\mp$ is proportional to $h^\dagger_{ee} h_{e\mu}$. This is tightly constrained by the $\mu \to eee$ data in Table~\ref{tab:limits}, leaving no hope to see any signal in this channel. In the $\tau$ lepton sector, the LFV decay constraints are comparatively much weaker. In the parameter space of interest $m_{H} \gtrsim m_{\tau}$, the limits on $|h^\dagger h|/m_{H}^2$ are almost constants, as in effective field theories with superheavy mediators. These constraints are all presented in Fig.~\ref{fig:p2}, with the shaded regions excluded.
The analytic formulae and calculation details are given in~\cite{supp}. 
As for the on-shell case above, the couplings $h_{e\beta}$ ($\beta = e,\, \mu,\, \tau$) are constrained respectively by the LEP $e^+ e^- \to \ell^+ \ell^-$ data. Thus we can set upper limits on the couplings $|h^\dagger_{ee} h_{e\tau}|$ and $|h^\dagger_{e\mu} h_{e\tau}|$, as shown in Table~\ref{tab:limits} and the left and right panels of Fig.~\ref{fig:p2}, which get weaker for lighter $H$, as in the on-shell case.

Given $h_{ee}$ and $h_{e\tau}$, the electron $g-2$ receives both contributions from the $H$ loops with an $e/\tau$ in the intermediate state, and for a fixed value of $|h_{ee}^\dagger h_{e\tau}|$, the weakest $(g-2)_e$ constraint occurs when $h_{ee} m_e \sim h_{e\tau} m_\tau$, with the two loops contributing almost equally. Similarly, one can obtain the $(g-2)_e$ limit on $|h_{e\mu}^\dagger h_{e\tau}|$, which induces the $\mu/\tau$-mediated diagrams. Both the constraints are presented in the left and right panels of Fig.~\ref{fig:p2}. Note that the muon $g-2$ can not be used to set unambiguous limits on the combinations $|h_{ee}^\dagger h_{\mu\tau}|$ and $|h_{e\mu}^\dagger h_{e\tau}|$, although the couplings $h_{\mu\tau}$ and $h_{e\mu}$ could contribute to  $(g-2)_\mu$ by themselves.

The dominant SM backgrounds are from the process  $e^+ e^- \to W^+ W^- \to e^- \tau^+ \bar{\nu}_e \nu_\tau$ which is expected to be small, if we require the two charged leptons to be back-to-back and their reconstructed energy $E_{\ell} \simeq \sqrt{s}/2$~\cite{Kabachenko:1997aw}. The angular distributions of charged leptons can also be used to suppressed the SM $WW$ backgrounds~\cite{Bian:2015zha}. Assuming 10 signal events as above, the coupling $|h_{ee}^\dagger h_{e\tau}|$ could be probed up to $6.5 \times 10^{-5} \, (6.0 \times 10^{-4})$ at CEPC (ILC) in the light $H$ limit, as shown in Fig.~\ref{fig:p2}. At the resonance $m_{H} \simeq \sqrt{s}$, the production cross section can be greatly enhanced by $m_{H}^2 / \Gamma_{H}^2$. To be specific, we have set the width $\Gamma_{H} = 10$ (30) GeV at $\sqrt{s} = 240$ GeV (1 TeV), where the prospects could be strengthened by roughly one order of magnitude (the dips in Fig.~\ref{fig:p2}). For $m_H>\sqrt{s}$, the production rate diminishes rapidly as $H$ becomes heavier. An off-shell $H$ could however be probed up to a few-TeV range, as shown in Fig.~\ref{fig:p2}, and ILC is expected to be more promising than CEPC in this mass range, as a result of the higher $\sqrt{s}$. 

The process $e^+ e^- \to \mu^\pm \tau^\mp$ could proceed via both the $s$ and $t$ channels, which depend on different couplings, namely $|h_{ee}^\dagger h_{\mu\tau}|$ and $|h_{e\mu}^\dagger h_{e\tau}|$, and are constrained respectively by the rare decays $\tau^- \to \mu^- e^+ e^-$ and $\tau^- \to \mu^+ e^- e^-$. 
Analogous to the $e\tau$ case above, a broad range of $m_{H}$ and $|h_{ee}^\dagger h_{\mu\tau}|$ could be probed in the $s$ channel, in particular in vicinity of the resonance, as shown by the middle panel of Fig.~\ref{fig:p2}. In the $t$ channel, the cross sections are comparatively smaller, and the detectable regions are much narrower, as shown by the right panel of Fig.~\ref{fig:p2}.

\begin{table}[b]
  \vspace{-13pt}
  \small
  \centering
  \caption[]{Reaches of the LFV couplings at future lepton colliders CEPC and ILC in both the on-shell and off-shell channel, with the BR of 1\%, 10\% and 100\% for $H$ decay reconstructible in the on-shell channel.}
  \label{tab:result}
  \begin{tabular}[t]{c|c|c|c|c}
  \hline\hline
  collider & BR & $|h_{e\mu}|$ & $|h_{e\tau}|$ & $|h_{\mu\tau}|$   \\ \hline
  \multirow{3}{*}{CEPC}
  & 1\% & [0.0026, 0.034] & $-$ & $-$ \\
  & 10\%  & [0.0099, 0.12] & [0.0009, 0.0096] & [0.017, 0.068] \\
  & 100\% & [0.00022, 0.050] & [0.00029, 0.015] & [0.0041, 0.10] \\ \hline
  \multirow{3}{*}{ILC}
  & 1\%  & [0.0099, 0.12] & $-$ & $-$ \\
  & 10\%  & [0.0014, 0.047] & [0.0056, 0.023] & $-$ \\
  & 100\%  & [0.00044, 0.050] & [0.00057, 0.054] & [0.046, 0.27] \\ \hline \hline
  \multicolumn{2}{c|}{ collider } & $|h_{ee}^\dagger h_{e\tau}|$ & $|h_{ee}^\dagger h_{\mu\tau}|$ & $|h_{e\mu}^\dagger h_{e\tau}|$ \\ \hline
  \multicolumn{2}{c|}{ CEPC } & $>6.5 \times 10^{-5}$ & $>1.1 \times 10^{-4}$ & $>2.0 \times 10^{-4}$ \\
  \multicolumn{2}{c|}{ ILC }  & $>6.0 \times 10^{-4}$ & $>1.0 \times 10^{-3}$ & $>1.5 \times 10^{-3}$ \\
  \hline\hline
  \end{tabular}
\end{table}

The future reaches of the LFV couplings in both the on-shell and off-shell production modes are collected in Table~\ref{tab:result}. It is clear that orders of magnitude of the couplings can be probed at future lepton colliders, i.e. from $\sim 10^{-4}$ up to ${\cal O} (0.1)$ for a scalar mass range of $\sim$ GeV to 200 GeV at CEPC (900 at ILC) in the on-shell channel, and couplings from $\sim \, 10^{-4}$ up to ${\cal O} (1)$ for a mass range from $\sim$ 100 GeV to few TeV in the off-shell mode. 

{\bf Conclusion.--}
We have shown that a hadrophobic neutral scalar $H$, which is well-motivated in a large class of new physics scenarios, can be probed in an $e^+e^-$ collider via its LFV couplings to the charged lepton sector. We present a model-independent analysis of how far the LFV coupling strengths and the scalar mass can be probed beyond the existing limits from the low-energy sector. In particular, we find that the full mass and coupling range of the scalar, that can explain the muon $g-2$ anomaly, can be tested in the future lepton colliders. This is largely complementary to the searches of LFV in the low-energy experiments and  hadron colliders.

\acknowledgments

{\bf Acknowledgements.--} The work of R.N.M. was supported by the US National Science Foundation under Grant No. PHY1620074.
Y.Z. is grateful to the Center for High Energy Physics, Peking University for the hospitality, the local support, and the active discussions during the visit.

\begin{widetext}
\begin{center}
\textbf{\large Supplemental Material}
\end{center}
\end{widetext}

\maketitle

\setcounter{equation}{0}
\setcounter{figure}{0}
\setcounter{table}{0}
\makeatletter
\renewcommand{\theequation}{S\arabic{equation}}
\renewcommand{\thefigure}{S\arabic{figure}}
\renewcommand{\bibnumfmt}[1]{[S#1]}
\renewcommand{\citenumfont}[1]{S#1}

\subsection{Example model frameworks}

{\it RPV SUSY:} A natural hadrophobic neutral BSM scalar appears in the minimal supersymmetric standard model with leptonic $R$-parity violation (RPV) in the form of the bosonic partners of the SM neutrinos, i.e. the sneutrinos $\tilde{\nu}$. The relevant RPV term is
\begin{eqnarray}
\label{eqn:LRPV}
{\cal L}_{\lambda} \ = \ \frac{1}{2}
\lambda_{\alpha\beta\gamma} \widehat{L}_\alpha \widehat{L}_\beta \widehat{E}_\gamma^c \,,
\label{eq:RPV}
\end{eqnarray}
with $\alpha$, $\beta$ and $\gamma$ flavor indices, $\widehat{L}$ and $\widehat{E}^c$ respectively the superfields with respect to the SM left-handed lepton doublets and the right-handed lepton singlets, and $\lambda$ the RPV coupling. The phenomenological consequences of Eq.~\eqref{eq:RPV} has been studied extensively, see e.g.~Refs.~[\blue{4}]. Given Eq.~(\ref{eqn:LRPV}), we can write explicitly the couplings in terms of the four-component Dirac spinors:
\begin{align}
{\cal L}_{\lambda} \ = & \ -\frac{1}{2}\lambda_{\alpha\beta\gamma} \Big(\tilde{\nu}_{\alpha L}\bar{\ell}_{\gamma R} \ell_{\beta L} + \tilde{\ell}_{\beta L}\bar{\ell}_{\gamma R} \nu_{\alpha L}  \nonumber \\
& \qquad \qquad   +\tilde{\ell}^*_{\gamma R}\bar{\nu}^c_{\alpha R} \ell_{\beta L} - (\alpha\leftrightarrow \beta) \Big) +{\rm H.c.} \, .
\label{eq:RPV2}
\end{align}
The first term in Eq.~\eqref{eq:RPV2} implies that the sneutrino couples to the charged leptons at the tree level in a LFV way if $\beta\neq \gamma$. Here either the CP-even or odd component of $\tilde{\nu}$ could be identified as the hadrophobic $H$  in Eq.~(1) with LFV couplings. There exist various constraints on the $\lambda$-couplings~[\blue{4}], but it is still possible to have some of the elements at the ${\cal O}(0.01-0.1)$ level, which could give rise to observable cLFV.

{\it Left-right symmetric model:} The minimal left-right symmetric model provides another natural framework to accommodate a hadrophobic neutral scalar in the form of the neutral component of the $SU(2)_R$-triplet scalar field $\Delta_R$. The Yukawa Lagrangian is given by
\begin{eqnarray}
{\cal L}_Y \ = & \ h_{q, \alpha\beta} \bar{Q}_{\alpha L} \Phi Q_{\beta R} +
\tilde{h}_{q, \alpha \beta} \bar{Q}_{\alpha L} \tilde\Phi Q_{\beta R} \nonumber \\
& \qquad + h_{\ell, \alpha\beta} \bar{\psi}_{\alpha L} \Phi \psi_{\beta R} +
\tilde{h}_{\ell, \alpha \beta} \bar{\psi}_{\alpha L} \tilde\Phi \psi_{\beta R} \nonumber \\
& \qquad  + f_{\alpha \beta} \psi^T_{\alpha R}Ci\sigma_2 \Delta_R \psi_{\beta R} +{\rm H.c.} \,,
\label{eq:LR1}
\end{eqnarray}
where $Q_{L,\, R}$ and $\psi_{L,\,R}$ are the left- and right-handed quark and lepton doublets respectively (with the heavy right-handed neutrino being the neutral component of $\psi_R$), $\Phi$ is the bidoublet scalar field, $\tilde\Phi = \sigma_2 \Phi^\ast \sigma_2$ with $\sigma_2$ the second Pauli matrix, and $h_\ell$, $\tilde{h}_\ell$, $f$ are independent Yukawa coupling matrices in the flavor space. After symmetry breaking, the CP-even neutral scalar sector of the model consists of three physical scalar fields, namely, $h$ (identified as the SM Higgs boson) and $H_1$ coming from the bidoublet scalar $\Phi$ and $H_3$ coming from the neutral component of the triplet scalar $\Delta_R$~[\blue{5}]. From Eq.~\eqref{eq:LR1} it is clear that at tree-level, $\Delta_R$ (and hence, $H_3$) does not couple to the SM quarks, and hence, is naturally hadrophobic. So $H_3$ can be identified as the hadrophobic scalar $H$ in Eq.~(1).

The LFV couplings of $H_3$ are induced from its mixing with the CP-even neutral components $h$ and $H_1$ of the bidoublet $\Phi$ through the quartic terms $\alpha_1{\rm Tr}(\Phi^\dagger \Phi){\rm Tr}(\Delta_R \Delta_R^\dagger)$ and $\alpha_2{\rm Tr}(\tilde\Phi^\dagger \Phi){\rm Tr}(\Delta_R \Delta_R^\dagger)+{\rm H.c.}$. The couplings of $H_1$ to the SM charged leptons can be written as $m_{D,\alpha\beta} H_1 \bar\ell_\alpha \ell_\beta / v_{\rm EW}$ with $m_D$ the $3\times3$ Dirac mass matrix for the type-I seesaw mechanism and $v_{\rm EW}$ the electroweak VEV. With the scalar mixing with $H_1$, the couplings of $H$ to the charged leptons are also proportional to the matrix $m_D$~[\blue{5}]. Thus, for large off-diagonal entries of $m_D$, this will give rise to observable cLFV effects. Such large off-diagonal elements can in principle be motivated from discrete flavor symmetries like $A_4$~[\blue{27}] or $Z_4$~[\blue{28}], which also explain why some of the other couplings can be zero (in the exact symmetry limit) or very small (generated by perturbations).

{\it Mirror model:} Another class of seesaw models for the tiny neutrino masses are the mirror models, where mirror leptons $\Psi_R$ are introduced, which are coupled to the SM leptons via a singlet scalar $\phi$:
\begin{eqnarray}
{\cal L}_Y \ = \
y_{\alpha \beta} \, \bar\psi_{\alpha L} \phi \Psi_{\beta R} + {\rm H.c.} \, .
\end{eqnarray}
Then the singlet scalar $\phi$ could couple to the SM charged leptons either at tree level through the mixing of the SM leptons with the mirror leptons, or at 1-loop level through the Yukawa interaction above and the trilinear scalar coupling, which arises naturally from the quartic terms in the scalar potential. With the flavor structure in the $y$ matrix, the effective coupling of $\phi$ to the charged leptons could be flavor-violating~[\blue{6}], thus providing another example of LFV hadrophobic scalar $H$.

{\it Two-Higgs doublet model:} There is a class of $Z_2$ symmetric two-Higgs doublet models (2HDM) called lepton-specific (or type-X) 2HDM, where one of the scalar doublets only couples to leptons:
\begin{align}
{\cal L}_Y \ = & \ Y^u_{\alpha \beta} \bar{Q}_{\alpha L} \tilde\Phi_2 u_{\beta R} + Y^d_{\alpha \beta} \bar{Q}_{\alpha L} \Phi_2 d_{\beta R} \nonumber \\
& \qquad + Y^\ell_{\alpha \beta} \bar{\psi}_{\alpha L} \Phi_1 \ell_{\beta R} +{\rm H.c.} \, .
\end{align}
In this case, both CP-even and odd neutral components of $\Phi_1$ are naturally hadrophobic and either of them (or a linear combination) can be identified as $H$ in Eq.~(1).
The LFV couplings of $H$ can be naturally induced by breaking the lepton-specific structure (e.g. with a soft mass term $m_{12}^2$), with the additional Yukawa couplings~[\blue{7}]
\begin{align}
\Delta {\cal L}_Y \ = & \ \xi^u_{\alpha \beta} \bar{Q}_{\alpha L} \tilde\Phi_2 u_{\beta R} + \xi^d_{\alpha \beta} \bar{Q}_{\alpha L} \Phi_2 d_{\beta R} \nonumber \\
& \qquad + \xi^\ell_{\alpha \beta} \bar{\psi}_{\alpha L} \Phi_1 \ell_{\beta R} +{\rm H.c.} \, .
\end{align}
In this case, the leptonic Yukawa couplings are proportional to $\left( \frac{m_{\ell_\alpha}}{v_{\rm EW}} \delta_{\alpha\beta} - \epsilon^\ell_{\alpha\beta} \right)$, where the non-diagonal coupling matrix $\epsilon^\ell$ is related to $\xi^\ell$ through a bi-unitary transformation. 
In light of the smallness of Yukawa couplings for the SM charged leptons, i.e. $m_{e,\, \mu,\, \tau} \ll v_{\rm EW}$, it is possible that the leptonic Yukawa couplings of $H$ are dominated by the $\epsilon_{\alpha\beta}$ terms, and some elements of $\epsilon$ might be much larger than the rest.

\subsection{Experimental constraints}

\begin{figure}[t!]
\includegraphics[width=7cm]{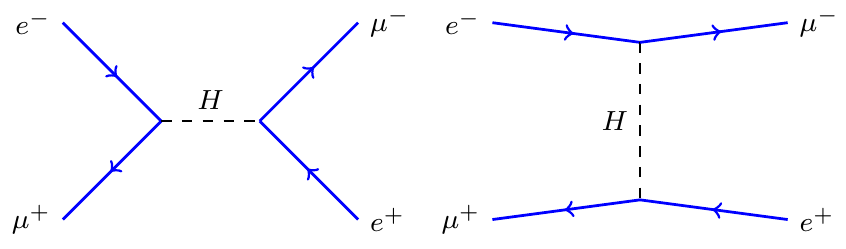}
\caption{Feynman diagrams for the $H$-mediated contribution to muonium oscillation.}
  \label{fig:diagram2}
\end{figure}

{\it Muonium-antimuonium oscillation:} The Feynman diagrams for the $H$-induced muonium oscillation are presented in Fig.~\ref{fig:diagram2}. The muonium-anti-muonium oscillation probability is given by~[\blue{29}]
\begin{eqnarray}
{\cal P} \ = \
\frac{2 |\Delta M|^2}{\Gamma_\mu^2 + 4|\Delta M|^2 }
\label{eq:prob}
\end{eqnarray}
with the $H$-induced mass splitting
\begin{eqnarray}
|\Delta M|  \ = \
\frac{2 \alpha_{\rm EM}^3 |h_{e\mu}|^2 \mu^3}{\pi m_{H}^2} \,,
\end{eqnarray}
with $\alpha_{\rm EM}$ the fine-structure constant, and $\mu = m_e m_\mu / (m_e + m_\mu)$ the effective mass. In the denominator of Eq.~\eqref{eq:prob}, the muon decay width $\Gamma_\mu$ is typically much larger than $2|\Delta M|$ for the range of mass and couplings considered in Fig.~3; therefore, ${\cal P} \propto |h_{e\mu}|^4/m_H^4$.

The di-muonium (not yet found experimentally), i.e. the bound state ${\rm Dm} \equiv (\mu^+ \mu^-)$, could also be affected by the LFV couplings in Eq.~(1), e.g. a $t$-channel $H$ could contribute to the decay ${\rm Dm} \to e^+ e^-$. Even if Dm is found in the low-energy experiments and its BR into $e^+ e^-$ is consistent with the SM prediction, the theoretical uncertainties would imply a bound $|h_{e\mu}|^2 / m_{H}^2 \lesssim {\rm GeV}^{-2}$~[\blue{30}], which is much weaker than the muonium oscillation constraint shown in Fig.~3.

{\it $\tau$ decay:} In the limit of $m_{H} \gg m_\tau \gg m_e$, the partial decay width of $\tau\to 3e$ is given by~[\blue{31}]
\begin{eqnarray}
\label{eqn:taudecay}
\Gamma (\tau^- \to e^+ e^- e^-) \ \simeq \
\frac{1}{\delta}
\frac{|h_{ee}^\dagger h_{e\tau}|^2 \, m_\tau^5}{3072 \pi^3 m_{H}^4} \,,
\end{eqnarray}
with the symmetry factor $\delta = 2$ accounting for the identical particles in the final state. The decay widths $\Gamma (\tau^- \to \mu^- e^+ e^-)$ and $\Gamma (\tau^- \to \mu^+ e^- e^-)$ are quite similar, with the couplings in Eq.~(\ref{eqn:taudecay}) replaced  by $|h_{ee}^\dagger h_{\mu\tau}|$ and $|h_{e\mu}^\dagger h_{e\tau}|$, and the symmetry factors of 1 and 2 respectively.

The presence of couplings $h_{ee}$ and $h_{e\tau}$ could induce the effective Wilson operators
\begin{eqnarray}
{\cal O}_{\tau e \gamma,\, (L,R)} \ = \
\frac{e}{8\pi^2} m_\tau
\left( \bar{e} \sigma^{\mu\nu} P_{L,R} \tau \right)
F_{\mu\nu}
\end{eqnarray}
with the effective coefficients, in the limit of $m_{H} \gg m_\tau \gg m_e$~[\blue{32}],
\begin{eqnarray}
\label{eqn:coeff}
c_{L} \ = \
c_{R} \ &\simeq& \
\frac{h_{ee}^\dagger h_{e\tau}}{8 m_\tau^6}
\bigg[ m_{\tau}^2 (2m_{H}^2 - m_\tau^2)  \nonumber \\
&& \left.
+ 2 m_{H}^2 (m_{H}^2 -m_\tau^2) \log \left( 1 -\frac{m_{\tau}^2}{m_{H}^2} \right) \right] \nonumber \\
\ &\simeq& \ \frac{h_{ee}^\dagger h_{e\tau}}{24 m_{H}^2} ~ + ~
{\cal O} \left(\frac{m_{\tau}^6}{m_{H}^6}\right) \,.
\end{eqnarray}
As expected, in the large $m_{H}$ limit, the effective coefficients are suppressed by $m_H^2$. Then the partial decay width
\begin{eqnarray}
\Gamma (\tau \to e \gamma)  \ = \
\frac{\alpha_{\rm EM} m_\tau^5}{64 \pi^4}
\left( |c_L|^2 + |c_R|^2 \right) \,.
\label{eqn:tau2decay}
\end{eqnarray}
The calculation of $\Gamma (\tau \to \mu \gamma)$ in presence of the couplings $h_{e\mu}$ and $h_{e\tau}$ is quite similar: in the limit of $m_{H} \gg m_\tau \gg m_{e,\,\mu}$ one has only to replace the couplings in Eq.~(\ref{eqn:coeff}) by $h_{e\mu}^\dagger h_{e\tau}$.

\begin{figure}[b!]
  \includegraphics[width=5cm]{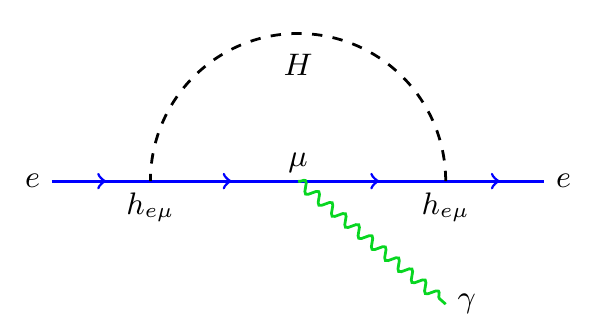}
  \caption{Feynman diagrams for the magnetic dipole moment of electron induced by $H$ with the coupling $h_{e\mu}$.}
  \label{fig:diagram3}
\end{figure}

{\it Lepton Magnetic Dipole Moments:} The Feynman diagram for the magnetic dipole moment of electron is shown in Fig.~\ref{fig:diagram3}, in presence of the coupling $h_{e\mu}$. The one-loop contribution is given by~[\blue{3}]
\begin{eqnarray}
\label{eqn:g-2}
\Delta a_e \ \simeq \
\frac{h_{e\mu}^2 m_e m_\mu}{16\pi^2 m_{H}^2}
\left[ 2 \log \left( \frac{m_{H}^2}{m_\mu^2} \right) - 3 \right] \,.
\end{eqnarray} It is trivial to get similar diagrams with the coupling $h_{e\tau}$ and also those for the muon $g-2$, by changing accordingly the flavors of the fermion lines and the couplings for the vertices. If the incoming electron  is replaced by $\tau$ and the internal $\mu$ line is replaced by $e$, then we have the LFV decay $\tau \to e\gamma$, which depends on the couplings $|h_{ee}^\dagger h_{e\tau}|$ (cf. Table I). In an analogous way we can have the diagram for $\tau \to \mu \gamma$ which has an electron mediator and depends on $|h_{e\mu}^\dagger h_{e\tau}|$.


\begin{thebibliography}{99}

\bibitem{PDG}
  C. Patrignani {\it et al}. (Particle Data Group),
  Chin. Phys. C, 40, 100001 (2016) and 2017 update.


\bibitem{deGouvea:2013zba}
  A.~de Gouvea and P.~Vogel,
  Prog.\ Part.\ Nucl.\ Phys.\  {\bf 71}, 75 (2013)
  [arXiv:1303.4097 [hep-ph]];
  H.~K.~Dreiner, K.~Nickel, F.~Staub and A.~Vicente,
  Phys.\ Rev.\ D {\bf 86}, 015003 (2012)
  [arXiv:1204.5925 [hep-ph]];
  A.~Vicente,
  Adv.\ High Energy Phys.\  {\bf 2015}, 686572 (2015)
  [arXiv:1503.08622 [hep-ph]];
  M.~Raidal {\it et al.},
  Eur.\ Phys.\ J.\ C {\bf 57}, 13 (2008)
  [arXiv:0801.1826 [hep-ph]].

\bibitem{Lindner:2016bgg}
  M.~Lindner, M.~Platscher and F.~S.~Queiroz,
  Phys.\ Rep.\  (2018)
  [arXiv:1610.06587 [hep-ph]].




\bibitem{susy}
C.~S.~Aulakh and R.~N.~Mohapatra,
  Phys.\ Lett.\  {\bf 119B}, 136 (1982);
  L.~J.~Hall and M.~Suzuki,
  Nucl.\ Phys.\ B {\bf 231}, 419 (1984);
  G.~G.~Ross and J.~W.~F.~Valle,
  Phys.\ Lett.\  {\bf 151B}, 375 (1985);
  R.~Barbier {\it et al.},
  Phys.\ Rept.\  {\bf 420}, 1 (2005)
  [hep-ph/0406039];



\bibitem{Dev:2016vle}
  P.~S.~B.~Dev, R.~N.~Mohapatra and Y.~Zhang,
  JHEP {\bf 1605}, 174 (2016)
  [arXiv:1602.05947 [hep-ph]];
  Phys.\ Rev.\ D {\bf 95}, no. 11, 115001 (2017)
  [arXiv:1612.09587 [hep-ph]];
  Nucl.\ Phys.\ B {\bf 923}, 179 (2017)
  [arXiv:1703.02471 [hep-ph]];
  A.~Maiezza, G.~Senjanovi\'{c} and J.~C.~Vasquez,
  Phys.\ Rev.\ D {\bf 95}, no. 9, 095004 (2017)
  [arXiv:1612.09146 [hep-ph]].









\bibitem{mirror}
  P.~Q.~Hung,
  Phys.\ Lett.\ B {\bf 649}, 275 (2007)
  [hep-ph/0612004];
  P.~Q.~Hung,
  Phys.\ Lett.\ B {\bf 659}, 585 (2008)
  [arXiv:0711.0733 [hep-ph]];
%
  J.~P.~Bu, Y.~Liao and J.~Y.~Liu,
  Phys.\ Lett.\ B {\bf 665}, 39 (2008)
  [arXiv:0802.3241 [hep-ph]];
  C.~F.~Chang, C.~H.~V.~Chang, C.~S.~Nugroho and T.~C.~Yuan,
  Nucl.\ Phys.\ B {\bf 910}, 293 (2016)
  [arXiv:1602.00680 [hep-ph]];
  P.~Q.~Hung, T.~Le, V.~Q.~Tran and T.~C.~Yuan,
  arXiv:1701.01761 [hep-ph].



\bibitem{2HDM}
  G.~C.~Branco, P.~M.~Ferreira, L.~Lavoura, M.~N.~Rebelo, M.~Sher and J.~P.~Silva,
  Phys.\ Rept.\  {\bf 516}, 1 (2012)
  [arXiv:1106.0034 [hep-ph]];
  A.~Crivellin, J.~Heeck and P.~Stoffer,
  Phys.\ Rev.\ Lett.\  {\bf 116}, no. 8, 081801 (2016)
  [arXiv:1507.07567 [hep-ph]].

\bibitem{supp} See {\it Supplemental Material} for more details of the model frameworks and experimental constraints of the LFV couplings, which includes Refs.~\cite{Kersten:2007vk, Dev:2013oxa, Clark:2003tv, Jentschura:1997tv, Sher:1991km, Harnik:2012pb}.


\bibitem{Kersten:2007vk}
  J.~Kersten and A.~Y.~Smirnov,
  Phys.\ Rev.\ D {\bf 76}, 073005 (2007)
  [arXiv:0705.3221 [hep-ph]].

\bibitem{Dev:2013oxa}
  C.~H.~Lee, P.~S.~B.~Dev and R.~N.~Mohapatra,
  Phys.\ Rev.\ D {\bf 88}, no. 9, 093010 (2013)
  [arXiv:1309.0774 [hep-ph]].

\bibitem{Clark:2003tv}
  T.~E.~Clark and S.~T.~Love,
  Mod.\ Phys.\ Lett.\ A {\bf 19}, 297 (2004)
  [hep-ph/0307264].

\bibitem{Jentschura:1997tv}
  U.~D.~Jentschura, G.~Soff, V.~G.~Ivanov and S.~G.~Karshenboim,
  Phys.\ Rev.\ A {\bf 56}, 4483 (1997)
  [physics/9706026];
  U.~D.~Jentschura, S.~G.~Karshenboim, V.~G.~Ivanov and G.~Soff,
  Phys.\ Lett.\ B {\bf 424}, 397 (1998)
  [hep-ph/9706401];
%
  I.~F.~Ginzburg, U.~D.~Jentschura, S.~G.~Karshenboim, F.~Krauss, V.~G.~Serbo and G.~Soff,
  Phys.\ Rev.\ C {\bf 58}, 3565 (1998)
  [hep-ph/9805375].




\bibitem{Sher:1991km}
  M.~Sher and Y.~Yuan,
  Phys.\ Rev.\ D {\bf 44}, 1461 (1991).

\bibitem{Harnik:2012pb}
  R.~Harnik, J.~Kopp and J.~Zupan,
  JHEP {\bf 1303}, 026 (2013)
  [arXiv:1209.1397 [hep-ph]].





\bibitem{CEPC-SPPCStudyGroup:2015csa}
  CEPC-SPPC Study Group,
  IHEP-CEPC-DR-2015-01, IHEP-TH-2015-01, IHEP-EP-2015-01.

\bibitem{Baer:2013cma}
  H.~Baer {\it et al.},
  arXiv:1306.6352 [hep-ph].

\bibitem{Gomez-Ceballos:2013zzn}
  M.~Bicer {\it et al.} [TLEP Design Study Working Group],
  JHEP {\bf 1401}, 164 (2014)
  [arXiv:1308.6176 [hep-ex]].

\bibitem{Battaglia:2004mw}
  E.~Accomando {\it et al.} [CLIC Physics Working Group],
  hep-ph/0412251.

\bibitem{Kabachenko:1997aw}
  V.~V.~Kabachenko and Y.~F.~Pirogov,
  Eur.\ Phys.\ J.\ C {\bf 4}, 525 (1998)
  [hep-ph/9709414];
  G.~C.~Cho and H.~Shimo,
  Mod.\ Phys.\ A {\bf 32}, no. 24, 1750127 (2017)
  [arXiv:1612.07476 [hep-ph]].



\bibitem{Ferreira:2006dg}
  P.~M.~Ferreira, R.~B.~Guedes and R.~Santos,
  Phys.\ Rev.\ D {\bf 75}, 055015 (2007)
  [hep-ph/0611222];
  J.~I.~Aranda, A.~Flores-Tlalpa, F.~Ramirez-Zavaleta, F.~J.~Tlachino, J.~J.~Toscano and E.~S.~Tututi,
  Phys.\ Rev.\ D {\bf 79}, 093009 (2009)
  [arXiv:0905.4767 [hep-ph]];
  B.~Murakami and T.~M.~P.~Tait,
  Phys.\ Rev.\ D {\bf 91}, 015002 (2015)
  [arXiv:1410.1485 [hep-ph]].



\bibitem{Banerjee:2016foh}
  S.~Banerjee, B.~Bhattacherjee, M.~Mitra and M.~Spannowsky,
  JHEP {\bf 1607}, 059 (2016)
  [arXiv:1603.05952 [hep-ph]];
  I.~Chakraborty, A.~Datta and A.~Kundu,
  J.\ Phys.\ G {\bf 43}, no. 12, 125001 (2016)
  [arXiv:1603.06681 [hep-ph]];
  I.~Chakraborty, S.~Mondal and B.~Mukhopadhyaya,
  arXiv:1709.08112 [hep-ph];
  Q.~Qin, Q.~Li, C.~D.~L\"{u}, F.~S.~Yu and S.~H.~Zhou,
  arXiv:1711.07243 [hep-ph].

\bibitem{Hays:2017ekz}
  C.~Hays, M.~Mitra, M.~Spannowsky and P.~Waite,
  JHEP {\bf 1705}, 014 (2017)
  [arXiv:1701.00870 [hep-ph]].

\bibitem{Rodejohann:2010bv}
  W.~Rodejohann and H.~Zhang,
  Phys.\ Rev.\ D {\bf 83}, 073005 (2011)
  [arXiv:1011.3606 [hep-ph]];
  T.~Nomura, H.~Okada and H.~Yokoya,
  arXiv:1702.03396 [hep-ph].

\bibitem{Calibbi:2017uvl}
  L.~Calibbi and G.~Signorelli,
  arXiv:1709.00294 [hep-ph].


\bibitem{Willmann:1998gd}
  L.~Willmann {\it et al.},
  Phys.\ Rev.\ Lett.\  {\bf 82}, 49 (1999)
  [hep-ex/9807011].







\bibitem{Mohr:2015ccw}
  P.~J.~Mohr, D.~B.~Newell and B.~N.~Taylor,
  Rev.\ Mod.\ Phys.\  {\bf 88}, no. 3, 035009 (2016)
  [arXiv:1507.07956 [physics.atom-ph]].

\bibitem{Hou:1995dg}
  W.~S.~Hou and G.~G.~Wong,
  Phys.\ Rev.\ D {\bf 53}, 1537 (1996)
  [hep-ph/9504311].

\bibitem{Abdallah:2005ph}
  J.~Abdallah {\it et al.} [DELPHI Collaboration],
  Eur.\ Phys.\ J.\ C {\bf 45}, 589 (2006)
  [hep-ex/0512012].



\bibitem{Belyaev:2012qa}
  A.~Belyaev, N.~D.~Christensen and A.~Pukhov,
  Comput.\ Phys.\ Commun.\  {\bf 184}, 1729 (2013)
  [arXiv:1207.6082 [hep-ph]].

\bibitem{Yu:2017mpx}
C.~Milstene, G.~Fisk and A.~Para,
  JINST {\bf 1}, P10003 (2006)
  [physics/0609018 [physics.ins-det]];
  D.~Yu, M.~Ruan, V.~Boudry and H.~Videau,
  Eur.\ Phys.\ J.\ C {\bf 77}, no. 9, 591 (2017)
  [arXiv:1701.07542 [physics.ins-det]].

\bibitem{Hammad:2016bng}
A.~Hammad, S.~Khalil and C.~S.~Un,
  Phys.\ Rev.\ D {\bf 95}, no. 5, 055028 (2017)
  [arXiv:1605.07567 [hep-ph]].



\bibitem{Bian:2015zha}
  L.~Bian, J.~Shu and Y.~Zhang,
  JHEP {\bf 1509}, 206 (2015)
  [arXiv:1507.02238 [hep-ph]].


\end{thebibliography}
\end{document}